\documentclass[12pt,a4paper]{article}

\usepackage{amsmath}
\usepackage{amsfonts}
\usepackage{amssymb}
\usepackage{amsthm}
\usepackage{slashed}

\newcommand{\SO}{{\mathrm{SO}}}
\newcommand{\ISO}{{\mathrm{ISO}}}
\newcommand{\DSU}{{\mathrm{DSU}}}

\newcommand{\dd}{{\mathrm d}}      % \d already defined
\newcommand{\D}{{\mathrm D}}

\newcommand{\R}{{\mathbb R}}   % Real numbers
\numberwithin{equation}{section}

\theoremstyle{definition}

\begin{document}

\title{Gauge gravity and discrete quantum models}

\author{John W. Barrett and Steven Kerr
%\thanks{}
\\ \\
School of Mathematical Sciences\\
University of Nottingham\\
University Park\\
Nottingham NG7 2RD, UK\\
\\
%E-mail john.barrett@nottingham.ac.uk
}

\date{}

\maketitle

\begin{abstract}

The gauge gravity action for general relativity    
in any dimension using a connection for the
Euclidean or Poincar\'e group and a symmetry-breaking scalar field is
written using a particularly simple matrix technique.          

A discrete version of the gauge gravity action for variables on a
triangulated 3-manifold is given and it is shown how, for a certain class of
triangulations of the three-sphere, the discrete quantum model this
defines is equivalent to the Ponzano-Regge model of quantum gravity.
\end{abstract}

%\maketitle

\newpage

\section{Introduction}
Analogies between the gravitational field and matter fields become important when one wishes to unify the two in a quantum model. One aspect of this is the relation between gravity and gauge fields: is gravity a special type of gauge theory \cite{S,K}, or are gauge theories an aspect of gravity \cite{Connes}?  The purpose of this paper is to shed some light on this debate by considering gravity in dimension $n$ as a gauge theory based on the Euclidean group $\ISO(n)$, or the Poincar\'e group, and to begin the construction of discrete quantum gravity models with this gauge symmetry.  

Early work on gauge gravity was done with the de Sitter or anti-de Sitter group, the simplest example being $\SO(n+1)$ in Euclidean signature. In the original Macdowell-Mansouri formulation \cite{MM}, the $\SO(n+1)$ gauge symmetry was explicitly broken to the tangent space group $\SO(n)$ in the action. Stelle and West \cite{SW} introduced an additional scalar field to carry out the symmetry breaking, so restoring the overall $\SO(n+1)$ symmetry when all fields are considered. This symmetry-breaking action was improved by Pagels \cite{Pagels} to give an action that is simpler for our purposes. A Poincar\'e group analogue of Pagels' action was developed by Grignani and Nardelli \cite{GN}.

The first part of the paper revisits this, presenting a simplified version in which the fields have a matrix representation that is close to Pagels' original $\SO(n+1)$ theory. Initially, Euclidean signature is chosen for simplicity, so that the gauge group is the Euclidean group $\ISO(n)$. In the action constructed here the spin connection and the frame field are packaged as parts of an $\ISO(n)$-valued connection form, and there is a scalar field with $\ISO(n)$ charge that effects a reduction in symmetry to $\SO(n)$, giving the usual first-order form of the Einstein-Hilbert action without a cosmological constant.  The symmetry breaking is the same mechanism as in the broken phase of a spontaneously-broken gauge theory. The scalar field can be thought of as a Higgs field that is constrained to lie in its vacuum manifold; the model does not have an unbroken phase.

The second part of the paper describes a model for three-dimensional space-times that is a discrete analogue of the functional integral similar to a state sum model, based on the $\ISO(3)$ gauge gravity action.  The construction depends on a certain geometric structure (a set of loops) for which we do not yet have a general definition that is valid for any topology of space-time. However it is possible to specify this structure, and hence the model precisely, for a certain set of triangulations of the three-sphere. It is shown that for these triangulations it reduces to the Ponzano-Regge model, with the structure specifying exactly the gauge-fixing required for the definition of the Ponzano-Regge model. There is a brief discussion of the extension of the model to a Lorentzian version, using as gauge group the Poincar\'e group in three dimensions. Finally the paper ends with some perspectives on the unsolved problems and possible future developments.

\section{Gravitational action}

First, Pagels' action  \cite{Pagels} for the gauge group  $\SO(n+1)$ is reviewed, with a brief explanation of the reduction to the usual Sciama-Kibble first-order action for general relativity when a suitable gauge is chosen. It is worth noting that the signature of the space-time and the sign of the cosmological constant can be changed by taking a different signature for the group, i.e., $\SO(n,1)$ or $\SO(n-1,2)$ instead of $\SO(n+1)$. 

Subsequently the analogous construction for the gauge group  ${\ISO}(n)$, the $n$-dimensional Euclidean group, is given. This naturally gives an action in which the cosmological constant $\Lambda$ is zero.

\subsection{$\SO(n+1)$ action - $\Lambda \neq 0$}

The Lie algebra for gauge group $\SO(n+1)$ can be represented as antisymmetric $(n+1) \times (n+1)$ matrices. Accordingly, the connection coefficients are one-forms $A^{BC}$, $B,C=1,\ldots,n+1$, with curvature components the two-forms ${F}^{BC}=\dd A^{BC} +A^{BD}\wedge A^{DC}$.
Indices are raised and lowered with the Euclidean metric $\delta_{AB}$. The symmetry-breaking is effected by a scalar field on space-time with values in a sphere in $\R^{n+1}$. This field has components $\phi^A$, with $\phi^A\phi_A=c^2$ and $c>0$ a constant, and the covariant derivative is
\begin{equation}\label{covariant}\D\phi^B=\dd\phi^B+{A^B}_C\phi^C.\end{equation}
Pagels' action is
\begin{align}\label{pagelsaction}
S = \int & (\D \phi)^A\wedge (\D \phi)^B\wedge\ldots 
\wedge  {F}^{XY} 
\epsilon_{AB\ldots XYZ} \phi^Z. 
% \label{nplus1action}
\end{align}
There are as many instances of  $\D \phi$ as is necessary in the action so that $\epsilon$ has $n+1$ indices.
 
For example, the four-dimensional action is 
 \begin{align}
S = \int & (\D \phi)^A 
\wedge  (\D \phi)^B 
\wedge {F}^{CD} 
\epsilon_{ABCDE} \phi^E. 
 \label{new action}
\end{align}

Let $N=n+1$. The $\phi$ field is a vector in $\R^N$ and so, by a gauge transformation, it may be rotated so that it points along the final coordinate axis,
\begin{align}\label{physicalgauge}
 \phi^Z \rightarrow \left( \begin{array}{c} 0 \\
 \vdots \\
 0 \\
 c \\
 \end{array} \right).
 \end{align}
This gauge choice is known as `physical gauge'. In this gauge, the remaining gauge symmetry is the set of transformations that fixes this vector, namely $\SO(n)$. To exhibit the fields in terms of $\SO(n)$ tensors, the vectors and matrices will be written in block form,
\begin{equation}\label{physicalgauge2}
 \phi^Z=\begin{pmatrix}0\\c\end{pmatrix}\end{equation}
with $0\in\R^n$, $c\in\R$, and
\begin{equation}\label{connection}
A^{BC} =\begin{pmatrix}\omega^{bc}&e^b\\-e^c&0\end{pmatrix}\end{equation}
 %\begin{align}\label{connection}
%A^{BC} = \left( \begin{array}{ccc|c}  & & &  e^1 \\
 %&  {\omega^{bc}} & &  e^2 \\ 
 %& & &  e^3 \\ \hline \rule{0pt}{2.4ex} %some extra vertical space
 %-e^1 & -e^2 & -e^3 & 0 \end{array} \right).
 %\end{align}
with an $n\times n$ matrix $\omega^{bc}$ and an $n$-dimensional vector of one-forms $e^b$.
 Capital indices $B,C=1\ldots N$ are in the fundamental representation of $\SO(N)$, and the corresponding lower case indices $b,c=1\ldots n$ are in the fundamental representation of the $\SO(n)$ subgroup.  Thus $A^{bc}=\omega^{bc}$, $A^{bN}=-A^{Nb}=e^b$, $A^{NN}=0$.

In the physical gauge, the 1-forms $e^b$ and $\omega^{bc}$ are interpreted as the components of the frame fields and spin connection respectively. Since $\phi$ is constant, $(\D\phi)^B = c A^{BN}$; the first $n$ components are given by $c e^b$ and the last component is zero. Defining $R^{ab}$ to be the curvature of the  $\SO(n)$-connection $\omega$, one has
\begin{equation} F^{ab}=R^{ab}- e^a\wedge e^b.\end{equation}
Thus the action becomes 
 \begin{align}\label{gravityaction}
 S = c^{n-1} \int  e^a  \wedge e^b\wedge\ldots\wedge R^{xy}\, \epsilon_{ab\ldots xy} - e^a \wedge e^b \wedge\ldots\wedge e^y\, \epsilon_{ab\ldots y} .
 \end{align}

This is the action for general relativity with a cosmological constant. The constant $c$ is related to the one dimensionless constant in pure gravity, namely the cosmological constant in Planck units. The action can be made to take its more familiar form by rescaling the frame field. Defining $\Lambda >0$ by $c^{n-1}\Lambda^{(n-2)/2}=1/G\hbar$ and setting $\tilde e^a=\Lambda^{-1/2}\,e^a$ results in the usual first-order form of the Einstein-Hilbert action (neglecting the numerical coefficients)
\begin{align}
 S = \frac1{G\hbar} \int  \tilde e^a  \wedge \tilde e^b\wedge\ldots\wedge R^{xy}\, \epsilon_{ab\ldots xy} -\Lambda \,\tilde e^a \wedge \tilde e^b \wedge\ldots\wedge \tilde e^y\, \epsilon_{ab\ldots y} .
 \end{align}

\subsection{${\ISO}(n)$  action - $\Lambda = 0$}

In this section the gauge gravity action for the Euclidean group is constructed by analogy with the action of Pagels. This uses a matrix representation of the gauge group that is rather similar to $\SO(n+1)$.  This results in an action in which the cosmological constant is naturally zero. When written in field components, the action coincides with the action studied in \cite{GN}.

\subsubsection{Representations of the Euclidean group}
The $n$-dimensional Euclidean group ${\ISO}(n)$ is the group of rotations and translations of $\R^n$. Its action is given by $x \rightarrow Mx + t$, where $M\in\SO(n)$ is the rotation matrix and $t\in\R^n$ is the translation vector.

This defining representation has a non-linear action but can be represented in terms of matrices if the dimension of the matrices is increased by one. The vector $x=(x^1,\ldots,x^n)$ is supplemented by an additional coordinate $x^{n+1}=c$. The action is
\begin{align}
\left( \begin{array}{c} x \\
c \\
\end{array} \right) &\rightarrow \left( \begin{array}{cc} M & t \\
0 & 1 \\
\end{array} \right) \left( \begin{array}{c} x \\
c \\
\end{array} \right) = \left( \begin{array}{c} Mx+ct \\
c \\
\end{array} \right). \label{vector rep}
\end{align}
This will be called the vector representation. Clearly, taking the non-linear subspace $x^{n+1}=c$ for a fixed constant $c$ gives the affine representation on $\R^n$, $x\mapsto Mx+ct$, which is analogous to the sphere in the previous section. The particular value $c=0$ gives a linear subspace that is a sub-representation
\begin{equation}\label{quotientrep}
\begin{pmatrix}x\\0\end{pmatrix}\rightarrow\begin{pmatrix} Mx\\0\end{pmatrix}
\end{equation}
in which only the rotations act.

The dual of the vector representation is called the covector representation. 
This may be represented as follows. Let $k=(k_1,\ldots,k_n)\in\R^n$ and $k_{n+1}=\Omega\in\R$ be the last coordinate. The action is
\begin{align}
\left( \begin{array}{c} k \\
\Omega \\
\end{array} \right) &\rightarrow \left( \begin{array}{cc} M & 0 \\
-M^{-1}t & 1 \\
\end{array} \right) \left( \begin{array}{c} k \\
\Omega \\
\end{array} \right)= 
 \left( \begin{array}{c} Mk \\
-\left(M^{-1}t\right).k + \Omega \\
\end{array} \right). \label{covector rep}
\end{align}
%This representation has a linear quotient representation \eqref{quotientrep} obtained by ignoring the last coordinate. 
The invariant contraction of a vector and covector is
\begin{align} x^Ak_A=x\cdot k + c\Omega=
 % \left( \begin{array}{cc} x & c  \\
%\end{array} \right) \left( \begin{array}{c} x \\
%c \\
%\end{array} \right) = 
 \left( \begin{array}{cc} Mx+ct & c  \\
\end{array} \right) \left( \begin{array}{c} Mk \\
-\left(M^{-1}t\right).k + \Omega \\
\end{array} \right) = x'\cdot k' + c\Omega' .
\end{align}
The primes here denote transformed quantities. 

The invariant bilinear form which can be used to contract two covectors is given by
\begin{align}
g^{AB} = \left( \begin{array}{cccc} 1 & & &  \\
&  \ddots & & \\
& & 1& \\
& & & 0 \end{array} \right), \label{upper metric}
\end{align}
since $ k'_A g^{AB} l'_{A} = k'\cdot l'=k\cdot l = k_{A} g^{AB} l_A $.

The invariant bilinear form which can be used to contract two vectors is given by
\begin{align}
g_{AB} = \left( \begin{array}{cccc} 0 & & &  \\
 & \ddots & & \\
& & 0 & \\
& & & 1  \end{array} \right) , \label{lower metric}
\end{align}
since $ x'^A g_{AB} y'^A = c^2 = x^A g_{AB} y^A $.
Finally, we note that the permutation symbols $\epsilon_{AB \ldots YZ}$ and $\epsilon^{AB\ldots YZ}$ are both invariant, since the transformations $\eqref{vector rep}$ and $\eqref{covector rep}$ both have determinant $1$.

\subsubsection{The action}
Now an $\ISO(n)$ invariant action will be constructed using these ingredients.  Since the bilinear forms \eqref{upper metric}, \eqref{lower metric} are degenerate, the operations of raising and lowering indices are not invertible and one has to be work out which quantities are naturally vectors or covectors. The analogue of the sphere for $\SO(n+1)$ is the subspace $x^{n+1}=c$ for the vector representation of $\ISO(n)$. This means that the field
$\phi^A = \left( \begin{array}{c} \phi^a \\
c \\
\end{array} \right) $
 is a multiplet of real scalar fields in the vector representation, with a fixed constant $c\in\R$ as the last component.

The action of an  $\ISO(n)$ connection on the vector representation is a matrix of one-forms
\begin{equation}{A^B}_C=\begin{pmatrix}{\omega^b}_c &e^b\\0&0\end{pmatrix}.\end{equation}
This equation is the analogue of \eqref{connection}. The covariant derivative \eqref{covariant} is
\begin{equation}\D\phi^B=\begin{pmatrix} \dd\phi^b +{\omega^b}_c\phi^c + ce^b\\0\end{pmatrix}.\end{equation}
Since the last component is zero, this lies in the sub-representation \eqref{quotientrep}, 
  transforming covariantly under rotations and not at all under translations.

Now consider the Euclidean field strength tensor, ${F}=dA + A\wedge A$. In terms of matrix components it is given by
\begin{align}
{F}^B_{\;\;\;C} = \left( \begin{array}{cc} \dd{\omega^b}_c + {\omega^b}_d \wedge {\omega^d}_c & \dd e^b + {\omega^b}_d \wedge e^d  \\
0 & 0  \end{array} \right).
\end{align}
Raising the second index using the metric $g^{AB}$ in $\eqref{upper metric}$ gives
\begin{align}
{F}^{BC} = {F}^B_{\;\;\;D} g^{DC} = \left( \begin{array}{cc} \dd{\omega^b}_c + {\omega^b}_d \wedge {\omega^d}_c  & 0  \\
0 & 0  \end{array} \right).
\end{align}
This tensor is also invariant under translations.

Finally an $\ISO(n)$ invariant action can now be constructed using the same formula as Pagels' action, \eqref{pagelsaction}.

This action can be gauge fixed, as in \eqref{physicalgauge}. The action reduces to
  \begin{align} \label{gravityaction2}
 S = c^{n-1} \int e^a  \wedge e^b\wedge\ldots\wedge R^{xy}\, \epsilon_{ab\ldots xy}  ,
 \end{align}
which is exactly the action for gravity \eqref{gravityaction} but without a cosmological constant. The action allows a rescaling of the frame field, which is equivalent to changing the value of $c$. 

It is worth noting that it is possible to add an independent cosmological term to this theory by starting with an additional term in the action
\begin{align}
 \int (D\phi)^A \wedge  (D\phi)^B \wedge\ldots\wedge (D\phi)^Y  \epsilon_{AB\ldots YZ} \phi^Z.
\end{align}

\section{Quantum ${\ISO}(3)$ model}

In this section, the ${\ISO}(3)$ theory will be quantised using a discrete formalism that is similar to a state sum model. The three-manifold is triangulated and the construction is a local formula involving variables associated to simplexes in the triangulation.

This construction is somewhat tentative and at present requires the manifold to have trivial topology. Since the classical $\ISO(3)$ action is equivalent to three-dimensional gravity after gauge-fixing, one expects similar results, and in fact our construction does give exactly the Ponzano-Regge model, and is therefore independent of the triangulation. The Ponzano-Regge model also requires some restrictions on the topology, and is regularised with some arbitrary choices that are equivalent to gauge-fixing, though the final result is independent of these choices. Thus the construction detailed here gives new angles on these aspects of the Ponzano-Regge model. In fact the construction of the Ponzano-Regge model starting from a classical action has always been only heuristic. Here the construction is somewhat more detailed as there is a discrete action as an intermediate step.

%However, here we quantise the theory in a different way that will remedy some unresolved issues in the Ponzano-Regge model. Namely, the quantisation will be formulated entirely on a triangulated manifold, with no need for the dual of the triangulation. The lattice action will therefore be fully gauge invariant, and without the introduction of superfluous variables connecting the triangulation and its dual. Further, the $\delta$-functions on group elements will be derived as an exact mathematical relation rather than being `heuristic', as is the case in the standard formulation of BF theory. 

\subsection{Discrete ${\ISO}(3)$ action}\label{discretesection}

A discrete analogue of the continuum Pagels action in three dimensions is proposed here. The construction will depend on a certain choice of a set of loops in the manifold; making this choice is postponed until a definite example is considered in section \ref{collapsiblesection}.

The continuum action \eqref{pagelsaction} on a closed $3$-manifold $M$ is
\begin{align}\label{3dpagelsaction}
S = \int (D\phi)^A \wedge {F}^{BC} \epsilon_{ABCD}\, \phi^D, %\label{action 1}
\end{align}
with the last component of field $\phi$ equal to a constant $c$,
\begin{equation}\label{lastcomponentc}\phi=\begin{pmatrix}\phi^1\\\phi^2\\\phi^3\\c\end{pmatrix}.\end{equation} 

This action has an additional local symmetry, besides the $\ISO(3)$ gauge symmetry that is built into the formalism \cite{GN}. This is particular to spacetime dimension three. The shift
\begin{equation}\label{shiftsymmetry}\phi\mapsto\phi+\psi,\end{equation}
 with
\begin{equation}\label{lastcomponent0}\psi=\begin{pmatrix}\psi^1\\\psi^2\\\psi^3\\0\end{pmatrix}\end{equation}
is a symmetry on a closed manifold. To prove this, note that $\Delta S = S(\phi+\psi)-S(\phi)$ has three terms, two of which are immediately zero. The third is  
\begin{align}
\Delta S = \int (D\psi)^A \wedge {F}^{BC} \epsilon_{ABCD}\, \phi^D,
\end{align}
which is also equal to zero after integrating by parts and using the Bianchi identity. This means that after gauge fixing the $\phi$ field to a constant, as in \eqref{physicalgauge}, the action still has $\ISO(3)$ gauge symmetry, not just $\SO(3)$ as one would expect from the analysis of the action \eqref{gravityaction2} in a generic dimension. This observation is in accord with the observation that three-dimensional gravity is an $\ISO(3)$ Chern-Simons gauge theory \cite{W-TCA}.

The discrete model is defined on a triangulation  $\Delta$ of $M$ and likewise depends on a constant $c$. The variables for the model are as follows.
\begin{itemize}\item For each oriented edge $e$ of the triangulation, an element $Q_e\in\ISO(3)$, such that for the opposite orientation $Q_{-e}=Q_e^{-1}$.
\item For each vertex $v$ of the triangulation, a vector $\phi_v\in\R^4$ with last component $c$, as in \eqref{lastcomponentc}. 
\end{itemize}

The edge $e$ has a starting vertex $s(e)$ and a finishing vertex $f(e)$, and the element $Q_e$ is interpreted as the parallel transport from $s(e)$ to $f(e)$.
The model is defined by choosing, for each $e$, an oriented loop of edges $\gamma(e)=e(n)\ldots e(2)e(1)$ that starts and ends at $f(e)$. These edges are arranged so that $f(e(k))=s(e(k+1))$, which means that the group elements can be composed to give a holonomy for the loop
\begin{equation} H(e)=Q_{e(n)}\ldots Q_{e(2)}Q_{e(1)}.\end{equation}

The question of how to choose the loop $\gamma(e)$ for each $e$ is an important one, but is postponed for now, with one stipulation. If the orientation of $e$ is reversed, there is a canonical choice
$\gamma(-e)=    (-e)  (- \gamma(e))e$, which will be assumed. This means that
\begin{equation} H(-e)=Q^{-1}_e H(e)^{-1} Q_e. \end{equation}

The discrete version of the integrand  of \eqref{3dpagelsaction} is given by an action for each edge $e$
\begin{equation}\label{discreteintegrand}L(e)=\left(\phi_{f(e)}-Q_e\phi_{s(e)}\right)^A\,\left(\log H(e)\right)^{BC}\,\epsilon_{ABCD}\,\phi^D_{f(e)}.\end{equation}
The first bracket is the obvious discrete version of the $\ISO(3)$ covariant derivative, and the second bracket the analogue of the curvature, with an index raised using $g$,
\begin{equation} \left(\log H(e)\right)^{BC}={\left(\log H(e)\right)^{B}}_F g^{FC},\end{equation}
and using the principal logarithm that maps the Lie group to a fundamental domain in the Lie algebra that has rotations with angle less than or equal to $\pi$. While this logarithm is not a continuous map, after integration the points of discontinuity on the boundary of the domain will be of measure zero and not cause any difficulty.

Due to the antisymmetry of the epsilon tensor, \eqref{discreteintegrand} can be simplified to
\begin{equation}\label{discreteintegrand2}L(e)=-\left(Q_e\,\phi_{s(e)}\right)^A\,\left(\log H(e)\right)^{BC}\,\epsilon_{ABCD}\,\phi^D_{f(e)},\end{equation}

The formula \eqref{discreteintegrand2} has the following important properties.
\begin{itemize}\item It is invariant under $\ISO(3)$ gauge symmetries.
\item It is independent of the orientation of the edge $e$.
\end{itemize}

The gauge transformations are the action of the elements $U_v\in\ISO(3)$ independently at each vertex $v$ and act on the connection variables as
\begin{equation} Q_e\mapsto U^{-1}_{f(e)}Q_eU_{s(e)}.\end{equation} 
The gauge-invariance follows from the fact that each of the indices $ABCD$ transforms in the vector representation of  $\ISO(3)$ acting at $f(e)$ and is invariant under the action of $\ISO(3)$ at all other vertices.

Changing the orientation of the edge $e$ gives
\begin{align}\label{reversediscreteintegrand}L(-e)&=-\left(Q_e^{-1}\,\phi_{f(e)}\right)^A\,\left(\log H(-e)\right)^{BC}\,\epsilon_{ABCD}\,\phi^D_{s(e)}\notag\\
&={\left(Q_e^{-1}\right)^A}_E\,\phi^E_{f(e)}\,{\left(Q_e^{-1}\right)^B}_F\,{\left(\log H(e)\right)^F}_G\,{Q_e^G}_I\,g^{IC}\epsilon_{ABCD}\phi^D_{s(e)}\notag\\
&=\phi^A_{f(e)}\,\left(\log H(e)\right)^{BC}\epsilon_{ABCD}\left(Q_e\phi_{s(e)}\right)^D\notag\\
&=L(e).
\end{align}

With these properties, it makes sense to define the action of the whole triangulated manifold as
\begin{equation}\label{discreteaction}S_\Delta= \sum_e L(e)\end{equation}
in which the sum is over the set of unoriented edges of  $\Delta$. This means each edge of the manifold appears just once in the sum, with an arbitrarily-chosen orientation used to define $L(e)$.

 This action is $\ISO(3)$ gauge-invariant. It does not, however, have the obvious analogue of the additional symmetry  \eqref{shiftsymmetry}. Pick a vertex $v$ and make the shift $\phi_v\mapsto \phi_v+\psi_v$, with the fields at all other vertices unchanged. The resulting change in the action is
\begin{equation}\Delta S_\Delta= -\,\psi_v^D\sum_{e\colon f(e)=v} \left(Q_e\phi_{s(e)}\right)^A \left(\log H(e)\right)^{BC}\epsilon_{ABCD},\end{equation}
summing over all oriented edges finishing at $v$. This formula follows by choosing the orientation of each edge $e$ that joins $v$ so that $v=f(e)$. There is no reason to believe that $\Delta S_\Delta$ will vanish in general.

\subsection{Gauge fixing}

In principle, one should define the partition function by integrating $\exp({iS_\Delta})$ over all the field variables. However both the gauge group and the space of the variable $\phi$ are non-compact and so integrating over gauge-equivalent configurations will give an infinite result. This can be circumvented, as usual, by integrating over the space of orbits of the gauge group. In practice, this is done by `gauge fixing': by a change of variables, the partition function can be written as an integral over some subset of the variables that are not coupled to the rest. These (infinite) integrals are then removed from the definition of the partition function. If these integrals are over a symmetry group, then the remaining action no longer has this symmetry.

By applying a suitable $\ISO(3)$ gauge transformation, the $\phi_v$ variables transform as $\phi_v\mapsto U_v\phi_v$, and so can all be taken to a constant vector
\begin{equation} \phi_v= \left( \begin{array}{c} 0 \\
0 \\
0 \\
c \end{array} \right).\end{equation}
This vector is fixed by $\SO(3)$ so the gauge-fixed theory is a discrete $\SO(3)$ gauge theory.

The Euclidean group element $Q$ can be decomposed as a product of a rotation $M\in\SO(3)$ and a translation $B\in\R^3$, i.e., $Q_e=B_eM_e$. 
 The action  \eqref{discreteaction} then simplifies to
\begin{align}
S_\Delta=  -c^2 \sum_{e} B_e^a  K_e^{bc} \epsilon_{abc},
\end{align}
with $K_e^{bc}=\log H(e)^{bc}$ and the indices $a,b,c=1,2,3$. These components of $H(e)$ only depend on the rotational part of the holonomy
\begin{equation} G(e)=M_{e(n)}\ldots M_{e(2)}M_{e(1)} \in\SO(3), \end{equation}
so one can write $K_e^{bc}=\log G(e)^{bc}$. 

With this gauge fixing, the formula for the partition function becomes, at least formally,
\begin{equation}\label{formalpartition} Z_\Delta=\int\left(\prod_e\alpha \dd B_e \dd M_e\right)\, e^{iS_\Delta}.\end{equation}
Here $\dd M$ denotes the Haar measure on $\SO(3)$ normalised so that $\int\dd M=1$,  $\dd B$ is the Lebesgue measure on $\R^3$ and $\alpha\in\R$ is a constant parameterising the possible normalisations of this measure.
There is still the problem that if the $K_e$ are not all independent then some of the $B_e$ integrals will be redundant and lead to an infinite result. This can be resolved by some further gauge fixing.

The equations of motion for this action, given by varying $B_e$ for each $e$, give immediately that $K_e=0$ and hence that the rotational holonomy of each loop $\gamma(e)$ is the identity. This action is very similar to the action for the Ponzano-Regge model, where the corresponding loops are of `dual edges' circulating each edge $e$. However in this model there are no dual edges and so it is necessary to find a new definition of the set of loops $\gamma(e)$. We do not know of a useful general definition for $\gamma(e)$, so this question is a challenge for further work. A definition of $\gamma(e)$ for a particular class of triangulated three-spheres, together with the necessary further gauge fixing, is given in the next section.

\subsection{Collapsible manifolds}\label{collapsiblesection}

This section describes a technical condition on a triangulation called collapsibility which is necessary for the construction of a concrete example of the discrete quantum gravity model.

A collapsing move on a simplicial complex $C$ can occur when there is a $k$-simplex $\sigma$ that is contained in only one $k+1$-simplex $\Sigma$. The move is the removal of both $\sigma$ and $\Sigma$ from the complex \cite{2cxbook}. The complex is said to be collapsible if it can be reduced to a point (a single vertex) by collapsing moves. In fact, if a complex is collapsible, one can always remove the simplexes in dimension order, i.e., in dimension three, remove all 3-2 dimensional pairs first, then 2-1 and finally the 1-0 pairs.

 For three-manifolds, consider a triangulated closed manifold $\Delta$ such that removing one tetrahedron $\tau_0$ results in a collapsible complex $C=\Delta-\tau_0$. Collapsibility puts very strong constraints on the topology, and in fact $\Delta$ has to be a triangulated three-sphere. 

The process of collapse is most easily described by introducing the dual vertices and dual edges.
% Note however that the dual triangulation is only introduced here out of convenience, and is not required for the definition of the model. Nonetheless it allows one to make contact with the Ponzano-Regge model more easily.
Given a triangulation $\Delta$, the dual vertices are placed at the barycentre of each tetrahedron in $\Delta$. Neighbouring dual vertices are connected by dual edges, which puncture the common triangle of their respective tetrahedra at its barycentre. 
%Given an edge $e$ in $\Delta$, the set of dual edges connecting the dual vertices belonging to the tetrahedra which $e$ is a part of form a polygon, which is the boundary of the dual face associated to that edge. And finally, given a vertex in $\Delta$, the set of dual faces associated to the edges which impinge on that vertex altogether form a polyhedral two-sphere, which is the boundary of the three-dimensional region dual to that vertex.

Collapsing the triangulated $3$-manifold $\Delta$ is described by a maximal tree $T$ of vertices and edges, and a second maximal tree $T^*$ of dual vertices and dual edges (called the dual maximal tree). First one removes all of the triangles which are punctured by $T^*$, and the interior of all tetrahedra. This can be envisioned as a `burrowing' process, where one removes $\tau_0$ and then burrows along the paths from $\tau_0$ defined by the dual maximal tree, removing all triangles and the interior of all tetrahedra that are encountered along the way.

The set of triangles, edges and vertices which remains after this is a simplicial $2$-complex denoted $\Delta_{T^*}$.  The middle stage of collapsing is to remove all the edges not contained in $T$ together with all the remaining triangles. In fact the duals of these form a third maximal tree of dual vertices and dual edges in the 2-complex $\Delta_{T^*}$, but this is uniquely determined by $T^*$ and $T$ and so is not independent data. After removing all the triangles, all that remains is $T$, which can be collapsed to any one of its vertices.

Notice that in the middle step the collapse sets up a $1-1$ correspondence between edges $e \notin T$ and triangles  $t\notin T^*$. Here a triangle is said not to belong to $T^*$ if it is not punctured by $T^*$. 
This map assigns to each edge $e \notin T$ the triangle $t(e)$ that it collapses through. Since $e$ is in the boundary of $t(e)$ there is also a natural correspondence between orientations of $e$ and $t(e)$.

\subsubsection{Partition function}

Using the above technical background of collapsing, it is now possible to specify exactly the partition function of the discrete quantum model for the triangulations of $S^3$ that are considered. The resulting model is the same as the Ponzano-Regge model for these triangulations.

The first step for defining the partition function for the collapsible examples is to specify the loops $\gamma(e)$. For $e\notin T$, the loop $\gamma(e)$ is defined to be the cycle of edges around the boundary of $t(e)$. This loop starts and ends at $f(e)$. Explicitly, if $\partial t(e)=e(1)+e(2)+e(3)$ and $e(3)e(2)e(1)$ is a composable path of edges with $e=e(3)$, then $\gamma(e)=e(3)e(2)e(1)$.

To complete the definition of the action \eqref{discreteaction} it is necessary to specify $\gamma(e)$ for $e\in T$. Here the simplest assumption is used, that $\gamma(e)$ is the trivial loop at $f(e)$ for these edges. The action  \eqref{discreteaction} is 
\begin{align}
S_{\Delta} = - c^2 \sum_{e \notin T}  B_e^a  K_{e}^{bc} \epsilon_{abc} = - c^2 \sum_{e \notin T}  B_e\cdot K_{e}.
\end{align}

The partition function of the theory is defined by integration; the $B_e$ variables for $e\in T$ do not appear in the action so the integrals over these variables are omitted. The result is the following definition for the partition function, which is now finite,
 \begin{align}
Z_\Delta &=   \int \left( \prod_{e \notin T } \alpha   \; \dd B_e \right) \;\left(\prod_e \dd M_{e}\right) e^{-i  c^2\sum_{e \notin T} B_e \cdot K_{e}} \notag\\
&=  \int \prod_e \dd M_{e}  \prod_{e \notin T }   \;   \frac{ (2\pi)^3 \alpha}{c^2}\delta^3(K_{e}). \label{partition function}
\end{align}
 The function $\delta^3(K)$  is related to the delta function at the identity on the Lie group $\SO(3)$ by 
\begin{align}
 \delta^3(K) = \frac{1}{2\pi^2} \delta(G) ,
\end{align}
with $K=\log G$, as follows from the fact that the measures are related by
\begin{align}
 dG = \frac{1}{2\pi^2}  \frac{\sin^2|K|}{|K|^2} \; \dd K. \label{measures}
\end{align} 
Hence if the constants are defined so that $c^2=4\pi \alpha$, then
\begin{align}\label{ponzanoregge}
Z_\Delta = \int  \prod_{e  } dM_e \prod_{e\notin T} \delta(G(e)) .
\end{align}
The holonomies $G(e)$ appearing in this formula are those around the triangles not in the dual tree $T^*$. Therefore $Z_\Delta$ is equal to the formula for the Ponzano-Regge partition function as defined in \cite{PR}. The formula has edges and dual edges interchanged but this makes no essential difference. 

The main point to make here is that not only is the partition function $Z_\Delta$ equal to the Ponzano-Regge model, but the local formula for the model is the same. This means also that the expectation values of the observables that are functions of the group elements, as defined in \cite{BH}, are the same in the two models. This gives invariants of the graph of edges of $\Delta$ known as the relativistic spin networks.

In fact as a particular special case, the partition function itself is equal to 1, as will now be explained. The rotational part of the connection can also be gauge fixed. This uses a gauge transformation to set the matrices $M_e\in\SO(3)$ to the identity for $e\in T$. 
The result is a formula similar to \eqref{ponzanoregge}, 
\begin{align}\label{ponzanoreggefixed}
Z_\Delta = \int  \prod_{e\notin T  } dM_e \prod_{e\notin T} \delta(G(e)) .
\end{align}
By changing variables in the integral this gives
\begin{align}\label{ponzanoreggefixed2}
Z_\Delta = \int  \prod_{e\notin T  } dM_e \prod_{e\notin T} \delta(M_e)=1 .
\end{align}
This formula is proved by induction on the number of triangles in the complex $C$. An inverse of a collapsing move on $C$ to make $C'$ adds one new variable, say $M_{e(1)}$, and one new delta function on a triangle, $\delta(M_{e(3)}M_{e(2)}M_{e(1)})$. The edges $e(2)$ and $e(3)$ belong to $C$ and so are either in $T$ or, by the induction hypothesis, their variables are constrained to the identity element by delta functions. Thus the part of \eqref{ponzanoreggefixed} due to the new triangle is
\begin{multline}\ldots\delta(M_{e(3)})\delta(M_{e(2)})\int\dd M_{e(1)}\; \delta(M_{e(3)}M_{e(2)}M_{e(1)})\\=\ldots\delta(M_{e(3)})\delta(M_{e(2)})\int \delta(M_{e(1)})\;\dd M_{e(1)}.\end{multline}

\subsubsection{ Quantum ${\ISO}^{+}(2,1)$ model}

The construction presented in the previous section is not a model of Lorentz\-ian quantum gravity in three dimensions because the gauge group is ${\ISO}(3)$, and yet there is a crucial factor of $i$ multiplying the exponent of the action in \eqref{formalpartition}. This situation can be remedied by instead using the group ${\ISO}^{+}(2,1)$, which is the semi-direct product of the connected piece of $\SO(2,1)$ with its Lie algebra. This gives a Lorentzian version of the previous construction.

The construction is the same as for $\ISO(3)$ except that the group is non-compact. The formula \eqref{discreteintegrand} is the same, using the logarithm for ${\ISO}^{+}(2,1)$ defined on a suitable fundamental domain in the Lie algebra.
Due to the infinite volume of the gauge group, the analogues of \eqref{partition function} and \eqref{ponzanoregge} are still infinite. However it is still possible that suitable observables are finite. Although the relativistic spin networks for $\SO^+(2,1)\times \SO^+(2,1)$ have not been studied, a similar issue arises with the relativistic spin networks for $\SO(3,1)$, which are finite for many graphs \cite{Baez:2001fh,Christensen:2005tr}. It is worth noting that in any case the $\ISO(2,1)$ model can be gauge fixed yielding a finite partition
function that is identical to \eqref{ponzanoreggefixed2}.

\section{Conclusion}
This paper gives a simple form of an action for general relativity in $n$-dimensions using as variables an $\ISO(n)$ connection.
We have begun to discuss the construction of discrete models similar to state sum models for quantum gravity based on this action, using as the first example a model in three dimensions that reduces to the well-known Ponzano-Regge model of quantum gravity.  It is already apparent that to generalise this example significantly some new structures are required. Geometrically, what is needed is some way of comparing edges and dual edges in a triangulation. This is similar to the framing issue in Chern-Simons theory, where the na\"ive construction of a quantum knot invariant needs the values of the gauge connection on the knot itself, which is exactly where it is singular \cite{wittenqft}. In fact the introduction of the collapsing of a manifold is something like framing data, and it would be interesting to understand this in a more general context.

Ultimately one would like to study higher-dimensional, and possibly higher-categorical, analogues of this construction, for which one hopes to find clues in the three-dimensional case. There is some interesting additional structure in the three-dimensional models that is worth mentioning here. As mentioned in section \ref{discretesection}, the gauge gravity action \eqref{3dpagelsaction} has an additional symmetry that does not survive in the discrete model.
Nevertheless one might wonder whether it can be incorporated in the discrete model in some way. It is well-known that the $\ISO(3)$ gauge symmetry of the Chern-Simons theory does not survive quantization; in fact the quantum symmetry is the corresponding quantum group, which is $\DSU(2)$ in this case \cite{BM}. This suggests there may be a role for $\DSU(2)$ symmetry in  discrete models similar to the one constructed here.
 
%%%%%%%%%%

\end{document}